\documentstyle[12pt]{article}

\def\NPB#1#2#3{Nucl. Phys. B {\bf#1} (19#2) #3}
\def\PLB#1#2#3{Phys. Lett. B {\bf#1} (19#2) #3}

\def\PRD#1#2#3{Phys. Rev. D {\bf#1} (19#2) #3}
\def\PRL#1#2#3{Phys. Rev. Lett. {\bf#1} (19#2) #3}

\def\preal{{\rm Re\,}}
\def\pim{{\rm Im\,}}

\def\s2{\frac{1}{\sqrt2}}

\def\beq{\begin{equation}}
\def\eeq{\end{equation}}
\def\beqa{\begin{eqnarray}}
\def\eeqa{\end{eqnarray}}
\def\IZ{\relax\ifmmode\hbox{\ss Z\kern-.4em Z}\else{\ss Z\kern-.4em Z}\fi}
\def\IP{\relax{\rm I\kern-.18em P}}
\def\IC{\relax\hbox{\kern.25em$\inbar\kern-.3em{\rm C}$}}

\def\cp#1{\relax\ifmmode {\IP\kern-2pt{}_{#1}}\else $\IP\kern-2pt{}_{#1}$\fi}

\def\-{\hphantom{-}}

\def\s2{\frac{1}{\sqrt2}}

\def\Tr{{\rm Tr \,}}
\def\IF{\relax{\rm I\kern-.18em F}}
\def\II{\relax{\rm I\kern-.18em I}}
\def\IP{\relax{\rm I\kern-.18em P}}
\def\IC{\relax\hbox{\kern.25em$\inbar\kern-.3em{\rm C}$}}
\def\IR{\relax{\rm I\kern-.18em R}}

\def\Dsl{\,\raise.15ex\hbox{/}\mkern-13.5mu D} 
\def\IZ{Z\kern-.4em  Z}
\def\cp#1{\relax\ifmmode {\IP\kern-2pt{}_{#1}}\else $\IP\kern-2pt{}_{#1}$\fi}

\topmargin
-1.5cm
\textwidth
15cm
\textheight
23.5cm
\oddsidemargin
1cm
\evensidemargin
1cm
\begin{document}

\makeatletter
\@addtoreset{equation}{section}
\makeatother
\renewcommand{\theequation}{\thesection.\arabic{equation}}
\pagestyle{empty}
\vspace*{1.0in}
\rightline{ IFT-UAM/CSIC-98-5; FTUAM-98/7}
\rightline{\tt hep-th/9804236}
\vspace{2.0cm}
\begin{center}
\LARGE{
New Perspectives in String Phenomenology from Dualities
\footnote{To appear in the proceedings of the
Workshop on Phenomenological Applications of String Theory (PAST),
ICTP, Trieste (Italy), October 1997.}
\\[10mm]}
\large{
Luis~E.~Ib\'a\~nez
\\[2mm]}
\small{
Departamento de F\'{\i}sica Te\'orica C-XI  \\[-0.3em]
and \\[-0.3em] 
Instituto de F\'{\i}sica Te\'orica C-XVI ,  \\[-0.3em]
Universidad Aut\'onoma de Madrid,\\[-0.3em]
Cantoblanco, 28049 Madrid, Spain. \\[-0.3em]
 }
\vspace{2.0cm}
\small{\bf Abstract} \\[5mm]
\end{center}

\begin{center}
\begin{minipage}[h]{14.0cm}
After a review of some topics concerning the 
phenomenological applications of perturbative string theory,
I discuss to what extent all of it is affected by the recent
developements in string dualities.

\end{minipage}
\end{center}
\newpage

\setcounter{page}{1}
\pagestyle{plain}
\renewcommand{\thefootnote}{\arabic{footnote}}
\setcounter{footnote}{0}

\section{Introduction}

The excitement of the theoretical physics community after the 
so called first string revolution in 1984 had in principle
a phenomenological basis. The cancellation of anomalies
for the $D=10$, $N=1$ superstring theories  with gauge symmetries
made in principle possible the idea of a unification of a
chiral gauge theory like the standard model (SM) with gravity
into a finite theory. String phenomenology 
\cite{quev} is just the
study in detail of that general idea: how is embedded the
SM (or perhaps the minimal supersymmetric standard model MSSM)
into string theory?  A good deal of effort has been devoted to this
field in the last ten years and some of its results have inspired
 other string theory areas as well as physics beyond the 
standard model in general. After the second 
(1995) string revolution 
\cite{sdual} 
(for reviews see \cite{sreviews} ) a natural question appears. How
is the general scheme of string phenomenology affected
by the discovery of the non-perturbative string dualities? 
It is possibly too soon to make any definite statement.
In spite of this, some qualitative features of the
new physics appearing in the new non-perturbative
and perturbative string vacua can already be extracted.

\section{\, Some general aspects of perturbative string phenomenology}

Circa 1985 the scheme for embedding the known
standard model (SM) interactions within the string scheme
were thought by many practitioners  to be relatively
simple and unique. If one starts from the $E_8\times E_8$ 
heterotic string and compactifies  on a Calabi-Yau
manifold one lands on an $N=1$  theory with gauge group 
$E_6\times E_8$. Such a construction gives rise to
a number of massless chiral $E_6$ generations given by
$n_g=\chi /2$, where $\chi $ is the Euler characteristic 
of the Calabi-Yau manifold \cite{phil}
. It was soon realized that
1) it is not clear whether  such an $E_6$ structure
is phenomenologically viable (problems with unwanted
extra massless matter, proton stability,
 neutrino masses, gauge coupling unification etc.);
2) there are many, many more ways to construct consistent
$N=1$, $D=4$ string vacua leading to a variety of gauge
groups (including directly the SM group) and massless
particle content; 3) the $SO(32)$ heterotic
is equally good from the point of view of model-building.
In the last ten years four-dimensional
string models  based on toroidal orbifolds, free-fermion
4-D strings or Gepner type of models have been constructed
\cite{phil,bert}.
Some of them  have a massless spectrum tantalizingly close to
the particle content of the minimal supersymmetric standard
model (MSSM). I think that this, by itself, is already an
achievement since they represent the first unified 
theories of all interactions including gravity.
Two broad classes of $D=4$, $N=1 $ string models have been
constructed, those which involve heterotic constructions 
where the algebra of the gauge group is realized at
Kac-Moody level $k=1$ and those with $k>1$. 
As we will recall below, the first class of models
(by far the most studied up to now) gives rise
to theories without adjoint Higgs fields and hence
one gets models with gauge groups like e.g.,
$SU(3)\times SU(2)\times U(1)^n$, $SU(4)\times SU(2)\times SU(2)$
or $SU(5)\times U(1)$  in which one can make the symmetry breaking
down to the SM without any adjoint Higgs. A number of three
generation models of these characteristics have been 
constructed  (see e.g., \cite{kinq,anton,ross,fara} ).
Models with $k>1$ have only been considered in the last five years 
or so
\cite{higher}. Three and four generation models with the massless 
spectra of $SU(5)$, $SO(10)$ or $E_6$ GUTs have been constructed.

In spite of the above successes, all of the
realistic  perturbative string vacua constructed up to now
have the general property of yielding extra unwanted massless
chiral fields beyond those present in the MSSM. One has to
abandon the string techniques and analyze the effective 
field theory. Then one has to assume that, after SUSY is broken,
some particular direction in the scalar field space is taken so that
(via Yukawa couplings) all of the unwanted massless particles
disappear from the low energy spectrum. Important phenomenological
properties like the quark-lepton spectrum and proton stability depend 
on the choice made for the pattern of gauge symmetry breaking.
That, of course, leads to a partial loss of predictivity.

There are a number of properties of the above 
perturbative heterotic vacua which are general and
appear in all different classes of constructions. 
They are particularly interesting because they may be
considered as generic predictions of perturbative string 
unification. Let us briefly recall some of these 
properties:

{\bf a)} In perturbative heterotic vacua the gauge coupling
constants are unified with the gravitational couplings:
\begin{equation}
G_{Newton}\ =\ {1\over {4 }}k_i \alpha _i^2 \alpha  '   
\label{unif}
\end{equation}
where $\alpha _i=g_i^2/4\pi $ $(k_i)$ is the coupling constant (KM level) of the
gauge group factor $G_i$
and $\alpha ' $ is the inverse string tension squared.
Thus in string theory one has 
unification of gauge coupling constants even in the
absence of a GUT group. The $k_i$ are integers $k_i\geq 1$ for
non-Abelian factors ($k_i=1$ in most models constructed) and 
fractional normalization factors for the $U(1)$'s.

{\bf b)} There is an upper bound on the rank of the gauge 
group in perturbative models. It comes from imposing the
cancellation of conformal anomalies (vanishing of the total central
charge, $c=0$ ) on the string world-sheet. It is easy to see that
it must be ${\rm rank \,} (G)\leq 22$. This 22 may be understood as coming
from the rank of the gauge group in $D=10$  before compactification
(${\rm rank \,} (E_8\times E_8)={\rm rank \,} (SO(32))=16)$  plus the maximum 
rank (six) of the gauge group one may obtain from Kaluza-Klein 
chiral $N=1$ compactification on the extra six dimensions.

{\bf c)}  There are general restrictions on the possible gauge
quantum numbers of massless fields in this class of theories.
For example, for an $SU(N)$ group realized at KM level $k$,
chiral fields which transform under the group as a representation
with Young-tableaux wider than $k$, cannot be present in the 
spectrum of the theory \cite{higher}. 
This implies, for example, that
models with adjoint Higgs fields (like GUTs) can only be
obtained for $k>1$.
 Furthermore, the (left-moving gauge ) conformal weight 
$h_{w}$ associated to a massless field must obey $h_w\leq 1$.
This implies that very large representations of the gauge 
group cannot possibly be in the massless spectrum of the theory,
since one can see that $h_w$ grows as the dimension of representation
grows \cite{bounds}. 
This is a very nice property of string theory since,
from the point of view of gauge field theories, there is no
reason at all to prefer lower dimensional representations
like those appearing in the SM.

{\bf d)}  There are no exact continuous global symmetries
in perturbative string theory
\cite{noglobal}. Whenever they seem to be present
they really correspond to local symmetries. This implies that
continuous global symmetries appearing at low energies 
(like e.g., baryon or lepton numbers) can only be approximate
(accidental) symmetries if seen from the string point of view. 

{\bf e)} Typically (although not always), when there are 
models with $U(1)$ gauge symmetries, there is one linear combination
which is apparently anomalous. In fact the anomaly is 
canceled by a four-dimensional version of the Green-Schwarz
mechanism. At the same time a dilaton-dependent Fayet-Iliopoulos
term proportional to the trace of the $U(1)$ charge over the 
massless spectrum is created.

{\bf f)} In all $D=4$, $N=1$ heterotic vacua there is a massless
complex chiral multiplet whose complex scalar field
(denoted $S$) contains the dilaton ($\preal  S=4\pi /g^2$). At the tree
level one has \cite{truncation}  :
\begin{equation}
f_{a}\ =\ k_aS \ \ ;\ \ K(S,S^*)\ =\ -\log(S+S^*)
\label{tree}
\end{equation}
where $f_a$ is the $N=1$ gauge kinetic function and 
$K(S,S^*)$ is the dilaton dependent piece  of the Kahler 
potential.

{\bf g)} $D=4$, $N=1$ string models with different 
numbers of chiral generations are (perturbatively) 
disconnected. The same is true for models with 
different numbers of supersymmetries. 

All the above are generic features of $D=4$, $N=1$ perturbative
string vacua. In addition a good amount of work on different
aspects of string phenomenology has been done in the last ten years.
Some of the directions particularly pursued in the literature are
the following \cite{quev}: 

{\bf i) } {\it Effective low-energy action}

A general $N=1$, $D=4$ supergravity Lagrangian is determined
(up to two derivatives)  by its
particle content and three functions of the scalars in the theory: 
 the kinetic function $f_a$ ($a$ runs over the different gauge groups),
the Kahler potential $K$   and the superpotential $W$. In principle
those functions can be perturbatively computed for any given $D=4$, $N=1$
heterotic vacuum. In practice this has only been done for 
some classes of Abelian toroidal orbifolds \cite{kahler}
and some fermionic models. Some information is
also known for some specific Calabi-Yau compactifications. 
In particular, the dependence of the one-loop (threshold)
corrections on the moduli ($T_i$) and complex structure ($U_j$)
scalar fields has been computed.
 These scalars characterize the size and shape
of the compactifying manifold. One finds for the Wilsonian
action a result of the form \cite{dkl}:
\begin{equation}
f_a\ =\ k_a\ S+\ \Delta _a(T_i, U_i)
\label{thres}
\end{equation}
The first piece is the tree level result which we already
mentioned above. The $\Delta _a$ piece is the 
moduli-dependent one-loop correction and is a holomorphic function.
 As I said it has been
computed for a large class of toroidal orbifold compactifications.
One of the most interesting properties found is that it has
definite properties under the $T-dualities$ of the underlying 
torus. In particular $\exp(\Delta _a(T_i) )$  behaves as a 
modular form with respect to the $SL(2,{\bf Z})$ symmetries
corresponding to those dualities.
The tree level Kahler potential and superpotential $W$ are also
known for some classes of orbifold and fermionic models
(see \cite{quev} for references).

{\bf ii)} {\it  Gauge coupling unification}

As we said, in string unification all gauge couplings 
meet at scale that should be close to the string scale
which has been computed \cite{kaplu}
to be of order $M_{string}=3\times 10^{17}$ GeV.
It is nice that this unification occurs but an $N=1$ 
extrapolation of low energy experimental results seems to indicate
a lower unification scale of order $2\times 10^{16}$ GeV.
Several  explanations have been proposed for this discrepancy. 
We direct the reader to ref.\cite{dienes} for a review and
references on the subject.

{\bf iii)} {\it SUSY-breaking and soft terms}

One of the phenomenological problems of $D=4$ string models 
(large rank of the gauge group) may be a virtue. These models
often contain ({\it hidden}) gauge interactions which do not
directly couple to the observable particles. If the gauginos
$\lambda _a$ of those hidden groups condense ($<\lambda _a\lambda _a>
\not= 0$) supersymmetry will in general be broken
\cite{gaugino}. Although this
is a nice possibility which may generate naturally a hierarchy of
scales, it is not free of problems. In particular, working
at the effective Lagrangian level one finds that the scalar 
potential has a qualitative dependence on the dilaton $S$ and
overall modulus $T$ of the form:
\begin{equation}
V(S, T)\ \propto \ {1\over {\preal  S \preal  T^3 }}
 |\ \exp{(3\preal  S)/2\beta }|^2    
\label{scalar}
\end{equation}
The exponential comes from the condensate (recall $\preal  S=4\pi/g^2$) and
$\beta $ is the one-loop beta function
of the confining gauge interaction. Here the scalar $T$ correspond to
the overall modulus which measures the size of the compact
manifold ($\preal  T=R^2$, $R$ being the compactification radius).
 From this equation one
concludes that the perturbative vacuum lies at $\preal  S\rightarrow \infty $
and/or $\preal  T\rightarrow \infty $ which correspond to a non-interacting 
and/or  a decompactification limit. These are the so called 
run-away problems. It has been argued \cite{sem} that the
large $T$ decompactification problem may disappear in some 
particular models if the one-loop threshold corrections
$\Delta _a(T_i)$  are taken into account. Indeed, if this is done
the above formula gets multiplied by a factor of the form
$1/|\Delta (T) |^6$ in such a way that two interesting things happen
\cite{sem}:
1) the scalar potential becomes invariant under the 
$SL(2,{\bf Z})$ modular invariance associated to $T$-duality.
2) An additional T-dependence appears in the potential
in such a way that for large $T$ the potential grows and the
$T$ vev is stabilized around the string scale  (one has to be careful
though to check that one remains in the perturbative regime).
A similar stabilization seems difficult for the $S$ field, at least 
if one remains within perturbation theory. In fact this possibility 
was one of the motivations in ref.\cite{sduality}  to introduce the concept
of $S$-duality in string theory. An interesting new proposal
for the stabilization of the dilaton field has been put forward in 
\cite{bmmq}.

Another approach to the problem of SUSY breaking at the level
of the effective Lagrangian was discussed in refs.\cite{soft,bims}   
(for a recent review see \cite{bimr}). The idea rests on the assumption that
SUSY is predominantly broken by the vevs of the auxiliary fields
$F_S$, $F_{T_i}$ of the dilaton/moduli fields present in large 
classes of $D=4$, $N=1$ heterotic vacua  (particularly in
toroidal orbifolds). This simple assumption (plus that of a
vanishing cosmological constant) leads to specific relationships
among the different soft terms possible in the effective
Lagrangian. Here is an example. Consider an $N=1$, $Z_N$ or
$Z_N\times Z_M$ heterotic orbifold. Any such a model
(it does not matter whether it is a $(2,2)$ or a $(0,2)$ 
compactification) has three sets of matter chiral fields 
associated to the three complex compact dimensions
(this is the untwisted matter sector of these models).
Then one can show that if SUSY is broken by 
an arbitrary combination of $F_S$ and $F_{T_i}$ ($i=1,2,3$)
the following relationship between SUSY-breaking
soft terms exist \cite{bims}:
\begin{equation}
m_1^2\ +\ m_2^2\ +\ m_3^2\ =\ M^2\ \ \ ;\ \ \ M\ =\ A_{123}
\label{soft}
\end{equation}
Here $m_i$ are the soft masses of the scalars in the three
untwisted sectors, $M$ is the gaugino mass and $A_{123}$
is the trilinear soft term associated to the Yukawa
coupling which relates the three types of
untwisted sectors. A particular case is that in which
only the dilaton auxiliary field $F_S\not=0$ (dilaton 
dominated limit) contributes. In that case one gets the
simple expression $\sqrt{3}m_i=M=-A$. Perhaps the most 
interesting aspect of this kind of relationships is that they
give rise to specific constraints on the supersymmetric 
spectra if applied to the MSSM. Thus one can hope to
test this kind of ideas if SUSY is found at LHC.
Another interesting point regarding these relationships
is their behaviour with respect to field theory finiteness.
Indeed, it turns out that boundary conditions for soft terms
of the type shown in eq.(\ref{soft}), when applied to $N=1$
two-loop finite theories, preserve finiteness \cite{finite}.

{\bf iv)}  {\it Anomalous U(1)'s }

Perturbative heterotic vacua in which there are gauged $U(1)$'s
often have one of those $U(1)$'s anomalous. In fact the anomaly
is canceled by a $D=4$ generalization of the Green-Schwarz 
mechanism \cite{gs} 
. This works as follows. The gauge kinetic term in the
Lagrangian is $k_aSW_aW_a$, where $W_a$ is the field strength 
superfield. Under an anomalous $U(1)_X$ gauge transformation
one has $\pim  S\rightarrow \pim  S - \delta _{GS}\Lambda (x) $,
 where $\Lambda (x)$ 
is the gauge $U(1)_X$ parameter and $\delta _{GS}$ is 
a constant model-dependent coefficient.
 Thus one can compensate 
an anomalous transformation from the standard triangle graphs
by an appropriate shift of the axion-like field $\pim  S$. For this to
work in a theory with gauge group
$U(1)_x\times \prod _a G_a$, $a=2,\cdots ,n$ the mixed anomalies
$A_a$, $a=1,\cdots, n$ of the $U(1)_X$ with all group factors must be
in the ratio of the KM level coefficients \cite{sin}: 
\begin{equation}
A_1\ :\ A_2\ : \cdots :\ A_n\ =\ k_1\ :\ k_2\ :\cdots :\ k_n
\label{gs}
\end{equation}
This has an interesting phenomenological consequence. Since the
$k_a$ give the normalization of the gauge coupling constants at the 
string scale, one can relate those normalizations to the mixed
anomalous of $U(1)_X$. Consider for example a situation in which
we have as part of the gauge group the SM one,
$SU(3)\times SU(2)\times U(1)_Y$. If there is an additional anomalous
$U(1)_X$ one obtains \cite{sin}: 
\begin{equation}
\sin^2\theta _W\ =\ {1\over {1+k_1/k_2}}\ =\ {1\over {1+A_1/A_2}}
\label{sin}
\end{equation}
where $A_1$, $A_2$ are the mixed anomalies of $U(1)_X$ with
$U(1)_Y$ and $SU(2)_L$ respectively. Thus the value of the
weak angle can be computed in terms of anomaly coefficients,
independently of any grand unification symmetry.
This possibility has been also used in the last few years to
construct models which predict definite patterns(textures) for
fermion mass matrices \cite{textu,ramond}.

The presence of an anomalous $U(1)_X$ has other dynamical
consequences. A (one-loop) dilaton dependent Fayet-Iliopoulos (FI) term 
proportional to $\Tr Q_X$ is also generated. In particular,
the  $U(1)_X$ D-term contribution to the scalar potential has the 
form \cite{fi}: 
\begin{equation}
V_X\ =\ {{g^2}\over 2} (\sum _i q_i|\phi _i |^2 + \Tr Q_X {g\over {192\pi ^2}}
)^2
\label{fi}
\end{equation}
where $\phi _i$ denotes the scalar fields which have charge $q_i$
with respect to $U(1)_X$. The presence of the second (FI) term
in this formula forces  some of the $\phi _i$ scalars to get
a non-vanishing vev. Thus the classical vacuum is unstable but
in all cases studied up to now there is a nearby supersymmetric
vacuum.

There are several other phenomenological aspects of
$D=4$, $N=1$ perturbative vacua with have also been studied
in the last few years including the possible role
of the $\pim  T_i$ fields as invisible axions, cosmological
constraints on the dilaton/moduli sector etc. We refer 
the reader to the reviews \cite{quev}.

\section{String theory dualities }

A lot has been learned about the non-perturbative structure of string
theories in the last three years or so. There is a good number of 
 reviews on the subject \cite{sreviews} 
and I will not try to describe here
these developements. Let me just describe the different connections 
which have emerged  among the different types of supersymmetric 
string theories. First of all, it was known that, upon toroidal
compactification of one dimension, Type IIA string is 
T-dual to IIB if we simply exchange the radius of 
compactification $R\rightarrow \alpha ' /R$ \cite{tdual}. 
This means that these two theories are perturbatively
equivalent. The same happens
with the two type of heterotic strings: $E_8\times E_8$ is
T-dual to the $SO(32)$ heterotic upon the same exchange
(in this case a particular Wilson line breaking both groups to the
common subgroup $SO(16)^2$ is also needed). These relationships
are perturbative in nature and were known since
 the mid-eighties \cite{tdual}.
New non-perturbative dualities have been found in the last three 
years \cite{sreviews} 
:  Type I , $SO(32)$ string theory is S-dual to the
$SO(32)$ heterotic  string. This means that the weak coupling of the
former is equivalent to the strong coupling limit of the latter.
Furthermore it has been found that Type IIA   theory and the
$E_8\times E_8$ heterotic may be both obtained from an underlying
11-dimensional theory termed M-theory. Type IIA string is obtained
from M-theory upon compactification on a circle, the string coupling
constant being given by the compactification radius. Thus when one
sends the 10-dimensional string coupling constant to infinity one 
recovers the eleventh dimension of M-theory. The $E_8\times E_8$
string is obtained upon compactification of M-theory on a segment.
The two $E_8$ factors are associated to the two boundaries of the
segment. Combining all these connections one obtains the 
remarkable result that all known $D=10$ string theories are different
perturbative limits of a unique underlying 11-dimensional 
structure, M-theory. Our knowledge of M-theory is still very limited
but we know that it contains membranes and fivebranes as fundamental
ingredients and that its low-energy limit gives rise to
11-dimensional supergravity.

These duality connections between  the different $D=10$ supersymmetric
strings gives rise upon compactification to lower dimensions to a
complicated network of non-trivial connections  between the different
theories. The apparent consistency of all the different derived 
connections in lower dimensions is one of the strongest arguments 
in favour of all this unique construction. For example, one
expects a duality between Type-IIA  compactified on a Calabi-Yau
manifold and heterotic $E_8\times E_8$ compactified on $K3\times T^2$
\cite{iu}.
Both compactifications yield $N=2$, $D=4$  theories and abundant
evidence has been found for this equivalence. New kinds of
perturbative and non-perturbative Type IIB vacua have also been constructed.
In particular a new non-perturbative way to obtain Type IIB
vacua has been termed F-theory
\cite{fth}. This theory lives in twelve dimensions,
although the extra two dimensions are dynamically frozen. 
Compactifications of this F-theory on complex  (elliptic)
Calabi-Yau n-folds yield $D=(12-2n)$-dimensional
theories with $N=1$ supersymmetry.
For example, F-theory compactified on a complex Calabi-Yau
four-fold is expected to be dual to heterotic compactifications
on Calabi-Yau 3-folds. Thus one expects to extract non-perturbative
information about heterotic vacua from the F-theory dual.

New {\it perturbative} Type I vacua have also been recently constructed.
This has renewed interest since its duality to the $SO(32)$
heterotic implies that one should be able to obtain non-perturbative
information on the latter by working with  Type I perturbative
vacua. In particular, one can  obtain Type I theory as a sort
of world-sheet parity orbifold ({\it orientifold}) of Type IIB
theory \cite{hor,orient} 
. In this construction of Type I,  the open strings appear as
the {\it twisted sector} of the Type IIB theory modded with respect to
world-sheet parity reversal (see below). New consistent
perturbative $N=1$, $D=4$ Type I vacua can be obtained 
\cite{fdorient,afiv} by  combining
the above orientifold action with discrete $Z_N$ twists analogous
to those appearing in standard heterotic orbifold models.

As we will discuss below, all this rich structure leads to a 
number of important implications for the possible phenomenological
applications of string theory. These implications can be classified as 
follows:

{\bf 1)}  New non-perturbative phenomena.

{\bf 2)} New classes of perturbative and non-perturbative
vacua .

{\bf 3) }  Extraction of non-perturbative information
about previously known vacua.

The study of all these aspects is still in its infancy 
so I will limit myself to discuss some expected general
new features in what follows.

\section{New general features}

{\bf a)} {\it Chirality-changing transitions}

Some of the new  phenomena appearing in $D=4$, $N=1$ string vacua 
may be more easily understood in terms of the simpler $D=6$, $N=1$
case. Let me then first review a few aspects of the latter type of
field theories \cite{iu}. 
In $N=1$, $D=6$ theories the relevant (non-gravitational)
SUSY multiplets are of three types: 1){\it Vector multiplets}. They
contain a vector field and its gaugino partner; 2) {\it Hypermultiplets}.
Contain a couple of complex scalars and their fermionic partners.
In general they transform under the gauge group of the vector multiplets
of the given theory; 3) {\it Tensor multiplets}. They contain a
two index antisymmetric field $B^{\mu  \nu }$, a real scalar $\phi $
and a fermionic partner. They do not carry quantum numbers with respect to the
gauge group.  Perhaps the simplest type of $D=6$, $N=1$ string vacua
may be obtained by compactifying the heterotic string on a Calabi-Yau
complex 2-fold (i.e., the K3 manifold). Consider for example the 
$E_8\times E_8$ heterotic. If we do the standard embedding of the
spin connection into the gauge connection we obtain a
$E_7\times E_8$ gauge symmetry and hypermultiplets
transforming like $10({\underline {56}})+65({\underline 1})$.
To obtain an anomaly free result one can check that
the background contains a total of 24 instantons.
This kind of perturbative vacua have {\it only one tensor}
 multiplet whose
real scalar $\phi $ is the string dilaton.
It is directly obtained after dimensional reduction of the
$D=10$ supergravity multiplet.
 Now, it was found  \cite{tensor} 
 that there can be non-perturbative transitions in
 the theory under which a particular (gauge anomaly free) combination
of 29 hypermultiplets transforms into a (singlet ) tensor multiplet:
\begin{equation}
1\ {\rm Tensor} \ \ \leftrightarrow \ \ {1\over 2} ({\underline {56}}) \ +\ 
({\underline 1}) 
\label{trans}
\end{equation}
This type of transitions occur when the size of one
the 24 underlying background instantons is set to zero. 
An analogue of this kind of transitions for the $D=4$, $N=1$ case is
also expected on general grounds \cite{chitrans,ebranes}. 
If we compactify the $E_8\times E_8$
heterotic on a Calabi-Yau  (complex three-fold)
 with standard gauge embedding we
get instead as gauge group $E_6\times E_8$  and we get 27-dimensional
chiral multiplets transforming like
$E_6$ fundamentals. Since we know that the $E_7$ fundamental branches as
${\underline {56}}\rightarrow {\underline {27}}+{\overline {\underline {27}}}
+2{\underline 1}$, non-perturbative 
(in general, chiral) transitions of the type
\begin{equation}
{\rm Singlets}\ \ \leftrightarrow {\underline {27}}
\label{transe}
\end{equation} 
are expected. More generally, anomaly-free combinations of charged 
chiral fields may be transmuted into chiral singlets and viceversa.
This applies, of course, to other groups and not only $E_6$.
This seems to imply that string vacua with different numbers of
e.g., chiral $E_6$ (or $SU(5)$, or SM) generations can be
non-perturbatively connected. This would  have far reaching 
consequences since it opens the way to a dynamical determination
of the number of quark-lepton generations. Indeed, in
perturbative string theory vacua with different numbers of chiral 
generations are disconnected and it seems difficult
a dynamical choice of the number of generations,  since the 
different vacua with different number of generations could not
be dynamically preferred. Of course, it is still not clear what
could eventually determine dynamically  the chiral particle spectrum.
Furthermore this kind of non-perturbative transitions in
$D=4$, $N=1$ have still to be better understood.

Apart from the above variation of the number of chiral
generations other possibilities appear to be (in principle)
opened. It is well known that both in GUTs or in string models
whenever there are extra unwanted particles in the massless 
spectrum there is essentially always the same idea to get rid
of them: look whether the model has appropriate Yukawa couplings so that,
by giving vevs to appropriate scalars, the unwanted fields become
superheavy. The new type of non-perturbative transitions seem to 
provide another mechanism by which one can get rid of unwanted charged
fields. An important difference is that whereas standard Yukawas
can only give masses to vector-like, non-chiral combinations of
 fields, the above non-perturbative transitions can also make to
disappear chiral (anomaly free)  collections of charged fields.
Let me emphasize however that these are  expectations which
still have to be realized in specific models.

{\bf b)} {\it Rank of the gauge group}

I already remarked how $D=4$, $N=1$ heterotic compactifications
have the rank of the gauge group bounded, ${\rm rank \,} G\leq 22$. It turns out
that non-perturbative effects can give rise to new gauge interactions
so that, in practice, the rank of the gauge group in string vacua 
is essentially unbounded above! Take again for a start the case of
heterotic vacua in $D=6$ with $N=1$ SUSY. Consider now the case
of the $SO(32)$ heterotic compactified on $K3$ with standard embedding in
the gauge group. The $SO(32)$ gauge symmetry is generically broken to
$SO(28)\times SU(2)$ and there are hypermultiplets transforming
as $10({\underline {28}}, {\underline 2})+65(1,1)$. As in the $E_8\times 
E_8$ case,  this corresponds to the presence of a background 
of 24 instantons (which turns out to be required for anomaly
cancellation). Each of these instantons have parameters which govern e.g.,
their size. It was found by Witten \cite{smallin} 
that if $n_I$ of those instantons
is put to zero size some interesting non-perturbative phenomena occur
(at finite coupling constant): 1) Additional gauge interactions
with the simplectic $Sp(n_I)$ gauge group appear and 2) Hypermultiplets
transforming as the antisymmetric representation of $Sp(n_I)$ and
$n_I$ vectorials of $SO(28)$ appear.  If the instantons are located
at some singularity of the $K3$ manifold, 
gauge groups $U(N)$ and $SO(2N)$ with a variety of hypermultiplet
representations may appear. In these more general cases 
extra tensor multiplets may also appear in the spectrum
\cite{iu}.
 Very singular configurations
of the instanton moduli space may yield very large gauge groups.
For example, if we consider a $E_8\times E_8$ compactification
on $K3$ and we locate all the 24 instantons at the same point and coinciding 
at a $K3$ singularity of the so called $E_8$ type, one gets
the gauge group \cite{philone}: 
\begin{equation}
E_8^{17}\times F_4^{16}\times G_2^{32}\times SU(2)^{32}
\label{cande}
\end{equation}
and 193 tensor multiplets!
The same kind of phenomena appear in $D=4$, $N=1$ non-perturbative
vacua. A particularly impressive gauge group is obtained
with certain heterotic compactification of $E_8\times E_8$
(whose dual may be obtained as an F-theory compactification
on certain Calabi-Yau four-fold). It yields a group \cite{philtwo}: 
\begin{equation}
E_8^{2561}\times F_4^{7576}\times G_2^{20168}\times SU(2)^{30200}
\label{tomaya}
\end{equation}
One must emphasize that these are {\it not typical} examples of gauge groups,
of course, well on the contrary they correspond to very, very particular
configurations in the moduli space of thousands of scalar fields.
But they certainly provide examples showing how the perturbative
heterotic bound ${\rm rank \,} G\leq 22$ is badly violated at the non-perturbative 
level. In fact one does not need to invoke non-perturbative effects to 
obtain large gauge groups (not so enormous as the above !)
in some particular (non-heterotic) string
constructions. They appear easily, as we will discuss below, in Type I
perturbative string constructions. However this fact was unknown
till the advent of the {\it D-brane} technology developed after
Polchinski's paper in 1995 \cite{polchi}.

{}From the phenomenological side the above fact suggests a number 
of comments. To start with, where should we embed the SM group?
It is all of it non-perturbative, perturbative or some
factor (e.g. QCD) is non-perturbative and the rest perturbative?
Due to the S-dualities probably it does not make sense to say that 
the SM group is fully non-perturbative or fully perturbative since
the whole idea of strong/weak coupling duality is the equivalence
of those two regimes (at least for relatively small gauge groups;
certainly there is no perturbative model yielding such extreme
spectrum as the ones above). But it could well be that part of 
the gauge group of the SM (or a GUT) or/and chiral particle content
could have non-perturbative origin. For example, the 
chiral multiplet spectrum of the MSSM  is suspiciously asymmetric:
quarks and leptons come in chiral representations whereas
the Higgs sector is vector-like. Perhaps the Higgs sector has
non-perturbative and the three generations perturbative origins
(or viceversa).

What seems clear is that the string perturbation theory used up to
now to explore string vacua misses most of the moduli space. We will
have to study the new possibilities opened up for embeddings
of the SM into string theory.

{\bf c)} {\it Multiple dilaton-like fields}

In perturbative $D=4$, $N=1$ heterotic vacua there is a single 
complex scalar field $S$ whose real part is the dilaton and which
couples in a universal manner to all gauge groups and matter fields.
In non-perturbative vacua there are in general more than one field
with similar characteristics. Let us again start with six dimensions.
Consider the compactification of the $E_8\times E_8$ heterotic on
$K3$ with standard gauge embedding. As we said, the perturbative
background must include 24 instantons, the gauge group is broken to
$E_7\times E_8$ and there are hypermultiplets transforming like
$10({\underline {56}})+65({\underline 1})$. This model has only
one $D=6$ tensor multiplet which includes one real scalar, the 
dilaton. This unique tensor multiplet comes from dimensional
reduction of the unique $D=10$ gravitational multiplet. We saw in the
previos section that in the $SO(32)$ case,
 when $n_I$ of the instantons in the background 
are put at the same point in the $K3$ and with zero size, 
some non-perturbative gauge group plus hypermultiplets appear.
When we do the same thing in the present $E_8\times E_8$
case something quite different happens 
\cite{iu}. What happens is that 
$n_I$ tensor multiplets appear  in the massless sector
(and some hypermultiplets disappear). This has a clear
M-theory interpretation which we refrain from explaining here
\cite{tensor}.
This is an example of a generic phenomenon. While 
perturbative heterotic vacua in $D=6$ have just one single
tensor multiplet which contains the dilaton, non-perturbative
vacua contain an indefinite number of such tensor multiplet.
The appearance of such type of tensor multiplets proliferates
(both in $E_8\times E_8$ and $SO(32)$ vacua)  if the 
zero size instantons are located at some $K3$ singularity.
We already mentioned that in the $D=6$ extreme example in previous
section there were 193 tensor multiplets! Each of them
contains one scalar with dilaton-like couplings to all the different gauge
groups in the theory. 
When compactifying further to $D=4$, $N=1$ theory  one
generically  gets from all
these tensor multiplets complex scalar fields $S_i$ with 
behaviour similar (though not identical)
to that of  the usual perturbative $S$ field. Furthermore
some of the dualities exchange the $T_i$ moduli fields of the
compactification with the dilaton. All in all, if the
string vacuum has a gauge group $\prod _a G_a$ one finds a 
general structure for the different gauge kinetic functions 
\begin{equation}
f_a\ =\ c_a^1S_1 +\ c_a^2S_2\ + \cdots +\ c_a^nS_n\ \  .
\label{fgauge}
\end{equation} 
where the $S_i$ represent here general massless scalars
with dilaton behaviour but may also include moduli type of fields
$T_i$ in specific cases. The $c_a^i$ are model dependent coefficients.
This implies that in general the gauge coupling constants of the
different gauge group factors are going to be different. 
Unlike perturbative heterotic vacua in which the tree level
$f$-function is proportional to a universal $S$ field, here things
are more complicated. Gauge coupling unification is no longer
automatic. Actually this is qualitatively not so different from the 
perturbative case: we already saw in eq.(\ref{thres}) that 
one-loop corrections to the $f$-function involve non-universal
gauge couplings for different gauge groups, but one may hope
those corrections to be smaller than the tree-level result. 
Here the dependence on the different $S_i$ fields of the $f_a$
functions is not expected to be particularly suppressed.
In particular, if one embeds the SM group directly 
into a string vacuum (without an intermediate GUT structure)
one runs into the risk of getting different boundary 
conditions for the three gauge coupling constants. This could
perhaps be an argument in favour of string GUT's
\cite{fphen}, although
one cannot exclude the existence of models in which
the different $SU(3)\times SU(2)\times U(1)$ couplings are
equal without a GUT symmetry (perturbative
models are an example).

Let us also point out that the proliferation of massless dilaton-like
fields makes also harder an analysis of soft terms along the lines
discussed in chapter 2.

{\bf d)} {\it Anomalous $U(1)$'s }

We already remarked that perturbative $D=4$, $N=1$ heterotic vacua 
often  contain {\it one} anomalous $U(1)_X$. The anomaly is
actually canceled by the four-dimensional version of the 
Green-Schwarz mechanism. Since there is only one
complex  dilaton field $S$ 
to do the trick,  there can only be one anomalous $U(1)_X$ and its
mixed anomaly with all the gauge groups have to obey eq.(\ref{gs}).)
In non-perturbative heterotic vacua (or in perturbative Type I
vacua) there are more than one field which can help 
in canceling $U(1)$ anomalies, there is a generalized 
$D=4$, $N=1$ Green-Schwarz mechanism
\footnote{  An analogous generalized mechanism was in fact known
to exist in $D=6$ \cite{sagnten} .}  at work
\cite{afiv}. There are two 
consequences of this : 1) there may be more than one anomalous 
$U(1)$ in the models  and 2) even in  cases in which 
there is a single anomalous $U(1)$, its mixed anomaly with respect
to the different group factors may be non-universal.
 One also expects the
presence of several Fayet-Iliopoulos terms depending on the different 
$S_i$ fields for each of the different anomalous $U(1)$'s.

The above facts may have implications on the
phenomenological  use of anomalous $U(1)$'s in order to
construct fermion mass matrices \cite{textu,ramond}. 
For example, one can consider new classes of models
with two anomalous $U(1)$'s or else models
with only one anomalous $U(1)$'s but non-universal
mixed anomalies with the different gauge groups etc.
Of course one loses in this case the simplicity of
the perturbative models.

\section{New $D=4$, $N=1$ String vacua }

Most of the new perturbative and non-perturbative string vacua
constructed using the new duality and D-brane techniques
are higher dimensional and with extended supersymmetry.
Much work remains to be made in the more 
phenomenologically interesting $D=4$, $N=1$ case.
Let me briefly review some of the latter constructions.

\subsection{F-theory compactifications on 4-folds}

This is not the place to present a review of F-theory
\cite{sreviews}. Let me
just describe a few main points. F-theory \cite{fth} is a new 
non-perturbative method to get consistent Type IIB vacua
in a variety of dimensions (including $D=4$, $N=1$ theories).
The bosonic massless fields in $D=10$  Type IIB theory 
include two scalars, the dilaton  $\phi $ (from the NS-NS sector) 
and a scalar $a$ (from the Ramond-Ramond sector). The particular 
 complex field  combination $\tau = a+ie^{-\phi /2}$ turns out to be
specially relevant. Indeed, Type IIB theory present a 
$SL(2,{\bf Z})$ S-duality which is generated by the modular
transformations in the $\tau $ field. The usual perturbative 
vacua take $\tau $ to be a constant. F-theory vacua allow $\tau $
to vary in a particular form consistent with the $SL(2,{\bf Z})$
S-duality of the theory. One identifies $\tau $ with the
complex structure of a torus $T^2$ living in some 
unphysical 11-th and 12-th dimensions. The idea is then to 
compactify this 12-dimensional F-theory on a manifold $M$
which {\it locally } looks like $M\propto T^2\times CY$
(i.e., an `elliptical fibration'), $T^2$ being in the 11-th and 
12-th dimensions. This gives rise to non-perturbative 
compactifications of IIB theory on the CY submanifold in the
fibration. 

In order to reach a $D=4$, $N=1$ theory one has to compactify
F-theory on a complex Calabi-Yau
four-fold which is an elliptic fibration \cite{fourfolds}.
The latter kind of manifolds are quite complicated and are not
so well known as the complex 3-fold Calabi-Yau 
manifolds abundantly used
for heterotic compactifications in the last fourteen years.
Furthermore extracting the spectra of these models
requires substantial expertise in algebraic geometry. Still  
some particular $D=4$, $N=1$ models have been studied,
although up to now the specific examples obtained seem to be
non-chiral. In spite of their complication, F-theory vacua 
are particularly well suited to extract non-perturbative 
information of string vacua. For example,  
superpotentials for some particular $D=4$, $N=1$ 
F-theory vacua have been obtained which have interesting 
modular behaviour under dualities \cite{superp}.

\subsection{ Brane cooking}

Dirichlet-branes (D-branes) have played a dominant role in many 
of the recent duality developements. Again, this is not the place 
to review this subject (see ref.\cite{branes}  for a nice pictorial
review). D-p-branes are solitonic states of Type II string theory
which have $p+1$ -dimensional worldvolume and carry Ramond-Ramond
charges \cite{polchi}. 
An important property of D-branes (which may serve as a
definition) is that open strings are allowed to end and  start on them.
D-branes have gauge fields living on their worldvolume and when 
sets of them are located at the same point they give rise to
non-Abelian symmetries like e.g. $SU(N)$ on the worldvolume.
Now, one can {\it cook} certain combinations of Type IIA
D-branes plus Neveu-Schwarz fivebranes distributed in such a way that
{\it on the worldvolume} of some particular D-branes lives
a $D=4$, $N=1$ theory of interest to us. This provides us
with a new geometrical tool to study some perturbative and
non-perturbative aspects of the {\it gauge theory
living on the worldvolume}. The simplest such 
configuration \cite{hannany}  corresponds to $N_c$ Type IIA D-fourbranes
stretching in between two Neveu-Schwarz fivebranes
\footnote{The latter are not, strictly speaking,  D-branes
but solitonic fivebranes associated to the 
the massless $B^{\mu \nu }$ field coming the  NS-NS sector
of the theory.}. On the 
worldvolume of the D-fourbranes lives a $D=4$, $N=2$ gauge
theory with $SU(N_c)$ gauge group.

Chiral models with $D=4$, $N=1$ can also be obtained by acting with some
orbifold actions
(see J. Lykken's contribution to these proceedings). 
Typically one gets gauge groups of the form
$U(N)^m$ with chiral matter in bifundamental representations of the
type:
\begin{equation}
(N, {\overline N}, 1,\cdots ,1) \ +\ (1,N, {\overline N},1,\cdots ,1)\ +\ \cdots 
\ + \ ({\overline N},\cdots ,1,N)
\label{bifund}
\end{equation}
Many other possibilities are however possible. One particularly nice
point about this type of constructions is that
they can often be embedded into M-theory and one can obtain
important non-perturbative information
 (like the Seiberg-Witten curves in the $N=2$ case).

It is important to remark that these kind of theories
are {\it not}  compactifications, gravity still lives in $D>4$
dimensions,
so they cannot be considered as they stand as $D=4$ unified
theories of all interactions. They are however an important tool to
study  the properties of pure gauge theories (possibly with matter) 
leaving on the worldvolume. It would be interesting to 
try to obtain in this way a brane configuration with an $N=1$
model living on the worldvolume resembling the MSSM. It will not
be a unified model of all interactions but perhaps could give us some 
interesting hints.

\subsection{Type IIB, $D=4$, $N=1$ orientifolds}

 These are  new classes of perturbative $N=1$ four-dimensional
strings whose structure was 
considerably  developed with the advent of the
D-brane technology in the last few years. Actually they may 
equivalently be considered as Type I vacua, as we will discuss
below. 

 Type I vacua 
with $D=4$, $N=1$ were essentially ignored before 1995
(see however ref.\cite{hor,orient} for early
work) . Type I
in $D=10$ has gauge group $SO(32)$,  and fully fledged 
four-dimensional strings based on it were  ignored because
of the technical difficulties in ensuring anomaly cancellation. The
latter has to be checked  case by case. On the other hand
the heterotic $SO(32)$ has also low energy group $SO(32)$ but
it is trivial to obtain general conditions for absence of anomalies in
$D=4$, $N=1$ models. Anomaly cancellation is a direct consequence of
world-sheet modular invariance of the closed heterotic string and it is
very easy to obtain general conditions on string (e.g., orbifold) 
backgrounds guaranteeing absence of anomalies. This is why no
serious attempt to do $D=4$, $N=1$ string model building based on Type I
was made before:  1) it is technically cumbersome to ensure 
absence of anomalies and  2) another $SO(32)$ (the heterotic) string
exists in which one can obtain equally interesting models with less effort !

There are a few reasons to reconsider Type I  vacua in the light of
recent developements. First of all, before 1995 
the existence and relevance of Dirichlet branes in the context
of Type I theory \cite{polchi} 
 was unknown. In fact their existence may be
considered as a consequence of T-dualities in the context of 
open strings. Secondly, as we remarked above, there is evidence of an
S-duality between Type I $SO(32)$ string theory and the 
$SO(32)$ heterotic. This means that strongly coupled heterotic
string is equivalent (dual) to weakly coupled Type I.  Thus one might
expect to obtain information about non-perturbative heterotic 
vacua by studying the perturbative (weakly coupled) Type I theory.
Actually, the strong and weakly coupled regimes get more
entangled when compactifying both theories to lower dimensions. 
Doing a dimensional reduction in both theories one finds that the
mapping between the dilatons in the different dimensions is
\cite{dil,apt} : 
\begin{equation}
\Phi _I\ =\ {{6-D}\over 4}\ \Phi _H \ -\ 
{{D-2}\over {16}}\ \log \det G_H^{(10-D)}
\label{dillower}
\end{equation}
where $D$ is the dimensionality of space-time and $\Phi _I$ $(\Phi _H)$
is the Type I (heterotic) dilaton (recall $\exp(\Phi )$ yields the
strength of e.g., gauge interactions in both theories). 
$G_H$ is the metric of the $(10-D)$ compact dimensions 
in the heterotic frame. Notice that for $D=10$, 
$\Phi _I=- \ \Phi _H$, so that indeed a strongly coupled heterotic 
corresponds to a weakly coupled Type I string. The case $D=6$
is special since the Type I dilaton is in fact mapped to the
overall modulus (compact variety size). Now in the phenomenologically
interesting $D=4$ case the coefficient of $\Phi _H$ is positive so that
there are different regimes. For moderate compactification radius
there is a weak$\leftrightarrow $weak coupling duality between 
Type I and heterotic in $D=4$. For large 
heterotic compactification radius there is a weak$\leftrightarrow $strong
duality. The two regimes can yield complementary information in
particular vacua \cite{apt}.

Let us discuss now in somewhat more detail how Type IIB $D=4$, $N=1$
orientifolds are constructed \cite{hor,orient,fdorient,afiv}. 
The idea is analogous to that of 
toroidal orbifolds in heterotic strings
\cite{orbif}. One starts from Type IIB
string and compactifies on $T^6/G$, where $G$ is some discrete
Abelian group ($Z_N$ or $Z_N\times Z_M$) which acts rotating the
lattice vectors defining the 6-torus $T^6$. This means that
there will be an `untwisted sector' in the spectrum obtained
by just a projection under $G$ of the $IIB/T^6$ spectrum and 
`twisted sectors' which correspond to IIB strings which are
closed modulo an element of $G$.
This is in general not
enough to get $N=1$ in $D=4$ since (unlike the heterotic case
which has only one) 
there are two $D=10$ supersymmetries in Type II strings
(one coming from right-moving closed string oscillations
and the other from the left-moving ones). 
To further reduce the number of supersymmetries one 
can mode the theory under the $Z_2$ operation termed 
{\it world-sheet parity} $\Omega $ \cite{hor}.  Let 
 $(\sigma , \tau )$ be the two (space-like and time-like)
worldsheet coordinates of the string. Defining the complex
world-sheet coordinate $z=\exp(\tau + i\sigma )$, one then has
$\Omega z={\overline z}$. Thus $\Omega $ transforms
left-moving  and right-moving vibrations of the string into each other,
so that the result of a projection of IIB string under $\Omega $
is a closed unoriented string with only one $D=10$ SUSY.
In addition it turns out that the consistency of the theory
requires the addition of twisted sectors with respect to 
$\Omega $. These are nothing but Type I open strings which have
to be added to the closed unoriented strings discussed above.
All in all this is a compactification of Type IIB
on $T^6/\{ G, \Omega \} $ yielding in fact a Type I
$D=4$ string theory with $N=1$.
 
Cancellation of anomalies is not guaranteed in this
$D=4$ models and has to be essentially imposed case by case
\cite{fdorient,afiv}.
This cancellation may be reinterpreted as the
vanishing of certain one-loop tadpole graphs. It is at this level
that the introduction of the D-branes is forced upon us.
For our purposes 
they may be defined as submanifolds of the full $D=10$ space
where the open strings can end. In the simplest type of
models we are describing here only D-ninebranes and D-fivebranes
turn out to be relevant. The world-volume of the ninebranes is the
full $D=10$ space. Thus the ends of open strings move freely
in all ten dimensions. There is an index $i$ attached to each
ninebrane as a label and open strings have then associated matrices
(the Chan-Paton matrices). 
For the case of the fivebranes the
worldvolume is some six-dimensional submanifold and they also
carry an index $j$ as a label. It turns out that for 
fivebranes to be present in this class of models the
orbifold discrete group $G$ must contain an order-two twist.
The fivebrane worldvolume spans the four uncompactified dimensions plus
one of the three compact complex planes. Thus
in this class of $D=4$ orientifolds there may be up to three
different sets of fivebranes depending on the particular 
complex plane occupied by their worldvolume. 
There will also  be  open strings stretching
between fivebranes which will also carry associated Chan-Paton
matrices.

Let me describe how is the general structure of the massless
spectrum of this kind of theories
\cite{fdorient,afiv}. As I said, they contain both
closed and open strings. From the closed strings one
gets the gravity multiplets as well as a number of 
moduli/dilaton singlet fields analogous to the $S$ and $T_i$ fields
of the heterotic. From the open strings one gets both gauge fields
and chiral multiplets transforming under them. Open strings
stretching among ninebranes give rise to some gauge group 
$G_9$ (typically with rank 16 or less). Each set of fivebranes 
yields some extra gauge group $G_{5_i}$, $i=1,2,3$. It is thus obvious that
the rank of the gauge group in Type IIB orientifold models 
exceeds in general the perturbative heterotic bound ${\rm rank \,} G\leq 22$.
There are chiral multiplets $C_9$ which are charged under $G_9$
and chiral multiplets $C_{5_i}$ which are charged with respect to the
corresponding $G_{5_i}$ group. In addition there are charged
chiral fields $C_{95_i}$, $C_{5_i5_j}$
 which transform simultaneously under different sets of groups. They come 
from open strings stretching between different classes of D-branes.

Let us show as a first example \cite{conformal} 
a model which only contains
fivebranes. In this example the Type IIB theory is 
compactified on $T^6$ and modded by
 the orientifold group
 $Z_6=Z_3\times Z_2$. Here $Z_3$ is the
standard $Z_3$ action in
 $D=4$ which involves $2\pi /3$ rotations on the
three compact  complex planes. The $Z_2$ is generated by $\Omega R$
(instead of simply $\Omega $),
where $R$ is a reflection of the first two complex coordinates. 
It turns out that
due to the fact that  $\Omega $ is not a generator of the
 orientifold group, there are no ninebranes.
The presence of the element $\Omega R$ and tadpole cancellation 
requires the presence of 32 Dirichlet fivebranes whose worldvolume 
lives in the four non-compact dimension plus the 3-d compact plane.
Now we will chose a particular configuration of the fivebranes 
on the fixed points under the $Z_3$  action which obeys
tadpole cancellation conditions. 
 Eight fivebranes will be sitting at 
fixed point at the origin.  The open
strings stretching among these fivebranes give rise to a  $U(4)$ group with
three 6-plets.  The remaining 24 fivebranes will be sitting at some other
fixed point away from the origin. The corresponding open strings give rise to
a gauge group $U(4)^3$ with chiral multiplets
\begin{equation}
3(4,{\bar 4}, 1)+3({\bar 4},1,4)+3(1,4,{\bar 4})
\label{finite}
\end{equation}
 The full chiral 
multiplet content (except for the dilaton field $S$ ) is shown in the table.
This model is chiral and has the interesting 
property that the  $SU(4)^3$ theory corresponding to the
fivebranes away from the origin is a finite theory \cite{conformal}.

\begin{table}
\begin{center}
\begin{tabular}{|c|c|c|c|c|c|}
\hline
$Sector $
& $SU(4)^4$ & $Q_x$ &  $Q_1$ & $Q_2$ & $Q_3$    \\
\hline
\hline
  Open Strings    & $3(1,{\overline 4},4,1)$  &   0  &   0   &  -1  &  1  \\
\hline
&  $3(4,1,{\overline 4}, 1, )$  &    0   &  1  &  0  &  -1  \\
\hline
&  $3({\overline 4}  , 4, 1, 1)$  &  
  0   & -1  &  1  &  0 \\
\hline
&  $3(1,1,1,6)$  & -2 &  0  &  0  &  0 \\
\hline
\hline
Closed   Twisted  Strings    & $27(1,1,1,1)$   &   0  &   0   &   0  &  0  \\
\hline
\hline
Closed   Untwisted  Strings    & $9(1,1,1,1)$   &   0  &   0   &   0  &  0  \\
\hline
\end{tabular}
\end{center}
\caption{ Chiral multiplets in the $Z_2\times Z_3$ Type IIB  orientifold.}
\label{t33}
\end{table}

The table also shows the charges of the different particles with respect to the
four $U(1)$'s. 
All of them are anomalous except for the linear combination
$Q_1+Q_2+Q_3$  which does not couple to any chiral multiplet.
In particular one 
finds $\Tr Q_X=-36$ (gravitational anomalies) and also the mixed anomaly
with respect to the last $SU(4)$ is $A_4=-6$ (the mixed anomaly with the
other three $SU(4)$'s vanishes).  This provides as with an example
of what we stated in the previous chapter:  in Type I vacua the mixed
anomalies of a $U(1)_X$ with respect to the different non-Abelian
factors are not equal. In the present model there is a generalized
Green-Schwarz mechanism in which some of the 27 twisted moduli 
participate. This model has a  simple {\it perturbative } 
heterotic dual which has 
a similar  (but not identical) massless particle spectrum.

Some of the simplest and more interesting types of $D=4$, $N=1$ orientifolds 
have both ninebranes and one set of
fivebranes whose worldvolume spans the four non-compact dimensions 
plus (say) the 3-d compact complex plane \cite{fdorient,afiv}.
 They are particularly interesting since,
having in general a gauge group with rank larger than 22, they can only
be dual to non-perturbative heterotic vacua. Thus the hope is to 
learn non-perturbative aspects of $D=4$, $N=1$ heterotic vacua by
studying the structure of their Type IIB orientifold duals. This class of
models have a gauge group with a structure
\begin{equation}
G_9\ \times G_5 \  .
\end{equation}
In these models both groups have (maximal) rank 16.
If the 
fivebranes are all located at the fixed point lying at the origin in the
first two complex dimensions, it turns out that one has $G_9=G_5$
and there is an explicit symmetry under the exchange of the 
spectra coming from ninebrane and fivebrane sectors. It turns out that this
symmetry is nothing but standard  (Type I)  
T-duality with respect to the
first two complex planes:  $R_i\leftrightarrow 1/R_i$, $i=1,2$. 
One can see that T-duality in Type I theory
exchange the roles of fivebranes
and ninebranes and the above fivebrane configuration
is self-dual with respect to T-duality.  
The untwisted closed string sector includes dilaton/moduli fields
with compactification radius dependence \cite{apt,afiv} 
\begin{equation}
S\ =\ e^{-\phi }R_1R_2R_3  + i\theta \ \ ;\ \ T_i=e^{-\phi }{{R_i}\over
{R_jR_k}} + i\eta _i  \ \ (i\not= j\not= k)
\label{dil}
\end{equation}
where $R_i$ is the size of the $i$-th complex plane.
These fields are the Type I duals of their well known
heterotic counterparts.
 From the open string sectors one gets three sets of chiral fields 
$C^9_i$, $i=1,2,3$ charged with respect to $G_9$, three sets of fields
$C^5_i$, $i=1,2,3$  charged with respect to $G_5$ 
and fields $C^{95}$ charged with 
respect to both. One finds the remarkable result that the
gauge kinetic functions of $G_9$ and $G_5$ are respectively:
\begin{equation}
f_9\ =\ S \ \ ;\ \ f_5\ =\ T_3 \  .
\label{ffuni}
\end{equation}
 This is to be compared to the heterotic perturbative result
eq.(\ref{tree}). Looked from the heterotic dual point of view, $G_9$ is a
perturbative gauge interaction whereas $G_5$ is non-perturbative.
Equation (\ref{ffuni}) shows that some non-perturbative
heterotic groups have gauge couplings governed by the compactification
moduli $T_i$ instead of the dilaton field $S$. Notice that under
T-duality in the first two complex planes one has \cite{afiv}

\begin{eqnarray}
R_1\leftrightarrow {{\alpha '}\over {R_1}}\  \ & :&\ \ 
  R_2\leftrightarrow {{\alpha '}\over {R_2}} \\
S &\leftrightarrow  &T_3 \nonumber\\ 
T_{1,2}&\leftrightarrow  &T_{2,1} \nonumber \\
C^9_i &\leftrightarrow  &C^5_i \nonumber \\
C^{95}&\leftrightarrow & C^{95} \nonumber 
\label{duales}
\end{eqnarray}
We observe that under T-duality the role of $S$ and $T_3$ are
exchanged. Looked from the heterotic dual side this 
exchange between the dilaton $S$ field and the modulus $T_3$ field
would look as a non-perturbative symmetry.

Typically, the gauge group in this kind of models has the form
\begin{equation}
G_9\ =\ G_5\ =\ U(M)\times U(N)\times U(P)\times  \cdots 
\label{grupo}
\end{equation}
whereas the chiral fields are usually bi-fundamental representations
such as e.g., $(M, {\overline N},1,\cdots ,1)$ 
$+(1,N,{\overline P},1,\cdots ,1)+\cdots $
etc., although there can also be some antisymmetric representations
among the $C^9_i$ and $C^5_i$ fields. This kind of representations
are of similar type to the ones corresponding to the 
left-handed quarks in the SM, which are $(3,2)$ representations
under $SU(3)\times SU(2)$. Notice,  however, that
Type I vacua can never give rise {\it at the perturbative level} to
spinorial representations of $SO(2N)$ groups nor exceptional groups
either (this is because in Type I theory the gauge group is
actually $SO(32)$ and not $Spin(32)$). Thus Type I theory is not the simplest way
to obtain e.g. $SO(10)$ GUT's (which have generations in
spinorial reps.) nor $E_6$ GUT'S (that have $SO(10)$ spinorial roots).

This kind of orientifold vacua have in general several anomalous 
$U(1)$'s \cite{afiv}. The anomaly is canceled by a generalized Green-Schwarz
mechanism in which not only $S$ but $T_3$ and other moduli fields
get involved (their imaginary parts get shifted under 
anomalous $U(1)$ transformations). This is an example of the 
phenomenon we described in section 4-d.

As we said these are models with gauge group $G_9\times G_5$ whose
rank may be as high as 32. If they are S-dual to some 
heterotic vacua,  certainly they must be  non-perturbative 
heterotic vacua. It is easy to find  
heterotic vacua ($SO(32)$ heterotic   on $Z_N$ or $Z_N\times Z_M$ 
orbifolds) whose  gauge group and untwisted charged fields 
precisely match the $G_9$ gauge group and $C^9_i$ orientifold massless
fields. However one finds that the candidate duals
in fact violate the perturbative modular invariance constraints
\cite{afiv}.
The heterotic duals of these class of orientifolds seem to
correspond to some sort of non-perturbative $SO(32)$
heterotic orbifolds. Heterotic orbifolds of this type
have been recently constructed in \cite{ebranes} 
 (see Aldazabal's contribution
to these proceedings).

Realistic $D=4$, $N=1$ Type IIB orientifolds have not yet been
constructed but it is certainly an interesting direction. 
The form of the Kahler potential   for some orientifolds of the type
discussed above have been obtained in ref.\cite{afiv} . The models 
seem to have quite a different phenomenology to that of
perturbative heterotic models in several respects, as the brief 
summary above shows. Furthermore they correspond by S-duality
to non-perturbative heterotic vacua and may perhaps teach us
some non-perturbative secrets of perturbative heterotic
models studied in the past. 

\subsection{M-theory compactifications on CY$\times S^1/{\bf Z_2} $ }

As we mentioned above, the strongly coupled limit of the
heterotic $E_8\times E_8$ string is M-theory compactified on a 
segment ($S^1/{\bf Z_2}$)
of length $\rho $. As $\rho \rightarrow 0$
one recovers the weakly-coupled heterotic string
\cite{hw}. The two boundaries 
of the segment correspond to the two $E_8$ factors of the 
heterotic which are purely ten-dimensional. 
Thus a way to obtain four-dimensional $N=1$ vacua corresponding
to strongly coupled heterotic models is to compactify 
M-theory on a Calabi-Yau$\times S^1/{\bf Z_2}$. In
principle this is true but our lack of sufficient knowledge
of the structure of M-theory  at the moment only allow us to 
extract some qualitative (but nevertheless interesting) features.
The low energy limit of M-theory is known to yield
11-dimensional supergravity. The massless sector of this theory is
particularly simple, it includes a graviton $G_{MN}$, a gravitino 
$\Psi _M$ and an antisymmetric tensor $C_{MNP}$. When compactified
on the segment $S^1/{\bf Z_2}$, $D=10$, $E_8$ super-Yang-Mills
fields are located at the two boundaries of the segment,
$X_{11}=0$ and $X_{11}=\rho $. The effective bosonic Lagrangian
involving gravity and $E_8$ fields has then the general form
\cite{hw,uniwitten}

\begin{equation}
L\ =\ - {1\over {2\kappa ^2 }} \int dX_{11} \sqrt{g} R\ -\ 
\sum _{i=1,2} {1\over {8\pi (4\pi \kappa ^2)^{2/3}} }\int dX_{10}
\sqrt{g} \Tr  F_i^2\ +\ \cdots 
\label{horwit}
\end{equation}
where $\kappa $ is the $D=11$ gravitational constant and the sum corresponds
to the two $E_8$ factors. Notice that the first integral extends over
the full $D=11$ space-time whereas the second is only ten-dimensional.
If we compactify six dimensions on a Calabi-Yau space of volume $V$
and the 11-th dimension on a segment of length $\rho $, one gets
a $D=4$ action
\begin{equation}
L_{4}\ =\ - {1\over {2\kappa ^2 }} V\rho \int dX_{11} \sqrt{g}  R\ -\ 
\sum _{i=1,2} {1\over {8\pi (4\pi \kappa ^2)^{2/3}} } V\int dX_{10}\ 
\sqrt{g} \Tr  F_i^2 \ +\ \cdots 
\label{horwitf}
\end{equation}
One can then identify the dependence of the 4-dimensional
gravitational Newton constant $G_N$ and gauge fine-structure
constant $\alpha _{GUT}$  on the different parameters and scales
\cite{uniwitten}: 
\begin{equation}
G_N\ =\ {{\kappa ^2}\over {16\pi ^2 V\rho }}\ \ ; \ \ 
\alpha _{GUT}\ =\ {{(4\pi \kappa ^2)^{2/3} }\over {2V} }
\label{constantsm}
\end{equation}
We will see in the next section how the dependence of $G_N$ on the
length $\rho $ may provide an interesting alternative to the
problem of gauge coupling unification in perturbative 
heterotic strings described in the second chapter. Doing a 
more careful dimensional reduction one can obtain
the qualitative behaviour of this strongly coupled
heterotic limit with respect to the equivalent to the
dilaton $S$ and overall modulus $T$ in this approach. One finds
a Kahler potential and gauge kinetic functions \cite{ovrut} 
\begin{eqnarray}
K\ & = &\ -\log(S+S^*\ -\ \epsilon |C|^2) \ -\ 3\log(T+T^*-|C|^2) \\
f_{1,2}\ & = & \ S\ \pm \ \epsilon T  \nonumber
\label{mkahler}
\end{eqnarray}
where the $C$ fields are charged matter scalars and
 the subindices $1,2$ correspond to the gauge groups 
$E_6$ and $E_8$ relevant for the standard gauge embedding.
$\epsilon $ is a model-dependent 
constant which is not expected to be particularly small.
Those familiar with effective Lagrangians in perturbative
string theories probably have noted how similar to the 
perturbative result the above 
 formulae are. 
There are a couple of differences though \cite{ovrut}:  1)
the $C$-dependent piece in the $S$-term of the Kahler potential
and 2) the $T$-dependent  pieces in the gauge kinetic functions.
Although similar corrections are in fact present 
\cite{in} at the one-loop 
level in perturbative vacua (see eq.(\ref{thres})), unlike
in that case,  here the coefficient $\epsilon$ may be of order one
\cite{nilyam} .
Furthermore,  the $S$ and $T$ scalar fields above
are defined as  \cite{ovrut} 
\begin{equation}
S\ =\ R^6\ +\ i \sigma \  + {{\epsilon}\over 2} |C|^2
\ \ ;\ \ T\ =\ \rho R^2 \ +\  i\eta \ +\ {1\over 2} |C|^2
\label{stm}
\end{equation}
where $R=V^{1/6}$ is the overall CY compactification radius
and $\sigma , \eta $ are axion-like fields. Notice that, 
whereas in perturbative heterotic vacua the size of the CY 
manifold was essentially given by $\preal  T$, in this non-perturbative
limit it is given by the $S$ field. Notice also 
that this non-perturbative 
exchange of the roles  of $S$ and $T$ appeared also when we discussed
a class of $D=4$, $N=1$ orientifolds.

The full meaning of these results and their applications to
different phenomenological questions like supersymmetry breaking
are at present being the subject of intense research \cite{mpheno}  by 
different groups and have been reported by several speakers at this
conference. I forward the patient reader to their contributions for
more details and complete lists of references.

\section{Gauge coupling unification: a hint of non
perturbative dynamics?}

I mentioned at the beginning of this talk the gauge unification 
problem of perturbative  heterotic unification. Let me now be a bit more
explicit. The gauge and gravitational bosonic terms in the
$D=10$ effective Lagrangian from the {\it perturbative} heterotic string
are
\begin{equation}
L_{10}\ =\ -\int dX_{10} \sqrt{g} e^{-2\phi } 
({4\over {\alpha '^4}}R\ +\ {1\over {{\alpha '}^3}}\Tr  F^2\ +\ \cdots  )
\label{phet}
\end{equation}
where $\phi $ is the dilaton field and $\alpha '$ is the inverse of the
string tension. Upon dimensional reduction to four dimensions via CY 
compactification on a manifold of volume $V$ one arrives at an effective
$D=4$ bosonic Lagrangian
\begin{equation}
L_{4}\ =\  -\int dX_{4} \sqrt{g} e^{-2\phi }V 
({4\over {\alpha '^4}}R\ +\ {1\over {{\alpha '}^3}}\Tr  F^2\ +\ \cdots ) .
\label{phetf}
\end{equation}
We then identify  Newton's constant and the gauge coupling 
\cite{uniwitten}: 
\begin{equation}
G_N\ =\ {{e^{2\phi }{\alpha '}^4}\over {64\pi V} }
\ \ ;\ \ \alpha _{GUT}\ =\  {{e^{2\phi }{\alpha '}^3}\over {16\pi V} }
\label{constf}
\end{equation}
Notice that one indeed has $G_N=\alpha _{GUT}\alpha '/4$ , as we remarked
in chapter 2. Consider now the mass of the massive gauge bosons in this
theory. The lightest (massive) ones will have a Kaluza-Klein mass
of order $M_{GUT}=1/V^{1/6}$$=(\pi\alpha_{GUT}^4)^{1/6}
/(4e^{2\phi }G_N^3)^{1/6}$.
Now, if we want to remain in the perturbative regime we have to imposse
$e^{2\phi }\leq 1$, which then implies $M_{GUT}^2\geq (\alpha_{GUT}^{4/3})
/G_N$. Plugging experimental numbers for $\alpha _{GUT}$ and
$G_N$ one gets a lower bound \cite{uniwitten} 
 $M_{GUT}\geq 10^{17}$ GeV,
an order of magnitud larger than the value obtained by
extrapolating low energy coupling data. 
More detailed analysis give results for the unification mass in perturbative
heterotic unification around a factor 20 larger than one would expect
on the basis of low energy data. This is the gauge coupling unification 
problem.

We already remarked that it is probably premature to say that this is 
indeed a problem and that several solutions have been put 
forward \cite{dienes}. 
But probably the most elegant one has been put forward
by Witten in ref.\cite{uniwitten}  . He has noted that lower unification
scales may naturally appear in the context of {\it strongly 
coupled} heterotic compactifications. Let us first consider the 
case of the $SO(32)$ heterotic string. Semi-realistic models may be
constructed with this heterotic string equally well as with the 
$E_8\times E_8$ heterotic.
 Now, we know that the strongly coupled limit 
of the $SO(32)$ heterotic is the {\it weakly coupled Type I}
string theory, whose gauge group in $D=10$ is also $SO(32)$. Let 
us now consider, as we did lines above, the effective low energy
bosonic Lagrangian but for the Type I case. It turns out that 
the Lagrangian in $D=10$ is identical to the heterotic case with
a particular difference: the gauge piece $F^2$ has an additional
$\exp(\phi )$ factor in front. When going down to four dimensions
this extra factor gives rise to a $G_N$ analogous to the 
heterotic one but $\alpha _{GUT}$ gets en extra factor $\exp(-\phi )$ 
compared to the  heterotic result and hence we get
\begin{equation}
G_N\ =\ {{e^{\phi }\alpha _{GUT}\alpha '}\over 4}
\label{newt}
\end{equation}
instead of the perturbative heterotic result 
$G_N=\alpha _{GUT}\alpha '/4$. Now we have an extra 
parameter to play and we can decouple the string scale 
$M_{string}=(\alpha ')^{-1/2}$ from the Planck scale $G_N^{-1/2}$.
In particular one can simultaneously have e.g. $\alpha _{GUT}=1/24$,
$M_{Planck}=10^{18}$ GeV and $M_{string}=10^{16}$ GeV with a
compactification scale $M_c\propto 4\times 10^{16}$ GeV.
All this is possible remaining in the Type I perturbative regime
($\exp(\phi )\propto 10^{-3}$). In this case one identifies 
$M_{GUT}$ with $M_{string}\propto 10^{16}$ GeV, in 
agreement with experiment. Thus within the context of
perturbative Type I vacua one can naturally evade the 
gauge coupling unification problem of perturbative 
heterotic strings. Since weakly coupled Type I string is
S-dual to strongly coupled  $SO(32)$ heterotic, one can
equivalently  say that one can solve this gauge
unification problem by going to the strongly coupled limit
of that heterotic string. 

The above discussion applied to models based on the
$SO(32)$ heterotic (or their Type I dual). What about $D=4$, $N=1$
models based on the $E_8\times E_8$ heterotic? We already 
recalled in the previous chapter that the strongly coupled limit 
of the $E_8\times E_8$ heterotic was M-theory compactified
on a CY$\times S^1/{\bf Z_2}$. We found above (eq.(\ref{constantsm})) 
the relationship
in this scheme between $G_N$, $\alpha _{GUT}$ and the
11-dimensional gravitational coupling $\kappa $, the CY volume $V$ and
$\rho $. Identifying the unification mass with the
CY Kaluza-Klein scale one obtains  $M_{GUT}=V^{-1/6}$$=(\alpha _{GUT}/
8\pi ^2 G_N^{2/3}\rho ^{2/3})^{1/2}$. Plugging the phenomenologically
appropriated  $G_N$, $\alpha _{GUT}$ values  one can get the 
correct GUT scale $M_{GUT}=10^{16}$ GeV by stting $1/\rho \propto 10^{14}$
GeV. This also fixes the fundamental M-theory scale $M_M={\kappa}^{-2/9}
\propto 2\times M_{GUT}$. Thus the overall structure of scales is as follows:
Around $10^{16}$ GeV one has the M-theory scale and only slightly below
one has the $M_{GUT}$ (or CY) scale. In between that scale and
$10^{14}$ GeV the world is five-dimensional but with some peculiar 
characteristics: gauge interactions and charged matter fields 
are purely four-dimensional and live at both boundaries ($X_{11}=0, \rho $)
of the 5-th dimension. On the other hand, gravitational fields, dilaton 
and moduli live in the bulk of the 5-dimensional space. Below
this intermediate scale $M_I=1/\rho \propto 10^{14}$ GeV there is a
$N=1$, $D=4$ field theory. This structure of mass scales may have
important consequences for different phenomenological aspects like
supersymmetry breaking and structure of SUSY breaking as well
as the possible role of moduli as invisible axions \cite{mpheno}.

All in all, the possibility of an underlying 
strongly coupled heterotic string provides an
atractive understanding of the unification of 
coupling constants within string theory. It is important to remark that
in both strongly coupled $E_8\times E_8$ and $SO(32)$ schemes
the GUT scale $M_{GUT}$ is a scale at which also extended
objets (M-theory in the first case, Type I strings in the second)
appear. Thus in both schemes one expects 
\cite{mpheno,caceres} the appearence of
operators of dimension $>4$ supressed by powers of $(1/M_{GUT})$ instead
of $(1/M_{Planck})$ as was previously thought for perturbative
heterotic unification. In particular, within the context of the MSSM one
would expect the generation of baryon number violating  dimension five 
operators like $[QQQL]_F$. If these operators are only supressed by
$(1/M_X)$, they would yield proton decay at rates excluded by present
proton stability limits. Thus in the schemes discussed above there must
be additional (e.g., discrete) symmetries which forbid this kind of
dangerous operators.

\section{Outlook}

Our improved understanding of string theory from the 
different duality connections,  give us both new insights
into possible new dynamics as well as new possibilities
for string vacua. There are new possibilities for 
obtaining potentially realistic $D=4$, $N=1$ string
models embedding and unifying with gravity the standard
model physics. Several  general properties of perturbative
heterotic unification which were thought to be general
predictions of string theory turn out  in fact to be predictions 
only of {\it perturbative } string theory. There may be
new non-perturbative 
transitions which change the number of chiral
generations. The gauge groups which can appear in string vacua 
may be quite large (for very particular configurations in the
moduli space of the scalars in each model). $D=4$,$N=1$ vacua 
may have in general more than one anomalous $U(1)$ and 
there may be more than one scalar field behaving like a dilaton.

The perturbative heterotic four-dimensional strings 
constructed in the last ten years constitute a particular class
of string vacua in which conformal field theory techniques are
applicable,  but new non-perturbative vacua exist which remain to be 
explored. In particular, Type II string theory which was essentially
ignored in the past for model building purposses, may give
rise to interesting $D=4$, $N=1$ vacua in terms of orientifolds or
F-theory techniques. Although
the underlying theory (M-theory) is unique, different
techniques may be more appropriate to study different
corners in the space of theories. For example, depending on the 
particular vacuum or coupling regime, the strongly coupled limit of 
a given perturbative $D=4$, $N=1$ heterotic vacuum may be reached in terms
of a Type II orientifold or/and F-theory compactified on a CY 
four-fold or/and M-theory compactified on $CY\times S^1/{\bf Z_2}$.
A systematic construction of this class of models is still to be 
done.

Appart from new vacua, the duality symmetries relate the strongly
coupled limit of some theories to the weak coupling limit of other
theories. This has been used in the last few years to extract
non-perturbative information for some classes of  $D=4$, $N=2$ 
vacua (like the exact prepotential). There is the hope to extend
the use of dualities to gain knowledge of non-perturbative
physics on some  $N=1$  (and perhaps $N=0$) vacua. This 
 could eventually give us a hint on how is broken the 
vaccum degeneracy (large number of  apparently consistent 
 perturbative string vacua).  The fact that M-theory is a unique
theory and all vacua seem non-perturbatively connected
give us hope that the dynamics should determine  things like
the number of chiral generations (hopefully  three!). In any event,
a new epoch starts for string phenomenology and it will take us some 
time to figure out the full structure of the new possibilities
which are open to us. The fact that gauge coupling unification
takes place at $10^{16}$ GeV more naturally within the context
of a strongly coupled heterotic string could be the first evidence 
in favour of a non-perturbative string unification of all interactions.

\bigskip

\bigskip

\bigskip


\centerline{\bf Acknowledgements}
I thank  G. Aldazabal, A. Font, D. L\"ust,
C. Mu\~noz,  A. Uranga and G. Violero
for discussions. I give special thanks to the organizers 
of this very stimulating meeting.

\end{document}